\documentclass[11pt]{article}
\usepackage{fullpage,citesort,epsfig,psfrag}
\usepackage[normal,small]{caption}
\newcommand{\beq}{\begin{equation}}
\newcommand{\eeq}{\end{equation}}
\newcommand{\bea}{\begin{eqnarray}}
\newcommand{\eea}{\end{eqnarray}}

\def\to{\rightarrow}
\newcommand{\be}{\begin{equation}}
\newcommand{\ee}{\end{equation}}
\newcommand{\bq}{\begin{eqnarray}}
\newcommand{\eq}{\end{eqnarray}}
\newcommand{\ket}[1]{|#1\rangle}
\newcommand{\bra}[1]{\langle#1|}

\def\ie{{\it i.e.\ }}
\def\m@th{\mathsurround=0pt }
\def\leftrightarrowfill{$\m@th \mathord\leftarrow \mkern-6mu \cleaders\hbox{$\mkern-2mu \mathord- \mkern-2mu$}\hfill
 \mkern-6mu \mathord\rightarrow$}
\def\overleftrightarrow#1{\vbox{\ialign{##\crcr
     \leftrightarrowfill\crcr\noalign{\kern-1pt\nointerlineskip}
     $\hfil\displaystyle{#1}\hfil$\crcr}}}

\begin{document}
\setlength{\captionmargin}{20pt}

\renewcommand{\thefootnote}{\fnsymbol{footnote}}
\begin{titlepage}
\begin{flushright}
UFIFT-HEP-02-14\\
hep-th/0204142
\end{flushright}

\vskip 3cm

\begin{center}
\begin{Large}
{\bf Liouville Perturbation Theory\footnote{
This work was supported in part by the Monell Foundation
and in part by the Department
of Energy under Grant No. DE-FG02-97ER-41029. 
}}
\end{Large}

\vskip 2cm
{\large 
 Charles B. Thorn\footnote{E-mail  address: {\tt thorn@phys.ufl.edu}}
}

\vskip 0.5cm
{\it School of Natural Sciences, Institute for Advanced Study,
Princeton NJ 08540}
\vskip0.20cm 
and
\vskip0.20cm
{\it Institute for Fundamental Theory\\
Department of Physics, University of Florida,
Gainesville FL 32611}


\vskip 1.0cm
\end{center}

\begin{abstract}\noindent
A comparison is made between proposals for the exact
three point function in Liouville quantum field theory
and the nonperturbative weak coupling expansion developed
long ago by Braaten, Curtright, Ghandour, and Thorn. Exact agreement
to the order calculated (\ie\ up to and including
corrections of order $O(g^{10})$) is found.
\end{abstract}
\vfill
\end{titlepage}
\section{Introduction}
\label{chap1}
Much progress has been made in our understanding of
Liouville quantum field theory since Polyakov
discovered its relevance to subcritical string theory
\cite{polyakovliouville}. In particular, bootstrap
ideas have led to proposals for the exact
three point correlators \cite{dorno,zam1995}
\bea
\langle e^{2\alpha_1\phi(x_1)}e^{2\alpha_2\phi(x_2)}
e^{2\alpha_3\phi(x_3)}\rangle =|x_{12}|^{2\gamma_3}
|x_{23}|^{2\gamma_1}|x_{31}|^{2\gamma_2}
C(\alpha_1,\alpha_2,\alpha_3)
\eea
of the exponentials of the Liouville field, whose dynamics is given
by the Lagrangian
\bea
{\cal L}=-{1\over4\pi}(\partial \phi)^2-\mu e^{2b\phi}.
\eea
Here the Liouville field theory is imagined to be defined on a two dimensional
sphere.

Many years ago, a systematic weak coupling expansion for
the Liouville theory, defined on a flat cylinder and
quantized using standard canonical methods, was developed
by Braaten, Curtright, Ghandour, and Thorn \cite{braatencgt}.
In this treatment the zero mode of the Liouville
field was treated exactly by its Liouville quantum
mechanics, but the nonzero modes were treated
perturbatively. In this way a weak coupling ($b\to0$) expansion 
to any finite order
could be developed for matrix elements of products of an arbitrary number
of operators $e^{2b\sigma_i\phi(x_i)}$ between energy
eigenstates. In particular, the matrix element
of $e^{2b\sigma\phi}$ between energy eigenstates with energies
of order $O(b^2)$ was expanded up to and including $O(b^{10})$ terms.
Since this matrix element is directly related to
the three point correlator for which an exact answer has
been proposed, we thought it would be useful to check
whether the exact proposal agreed with the much older
weak coupling calculation. In this brief note
we show that there is exact detailed agreement to the
order calculated. This is important, first because it
gives another piece of evidence in favor of the 
proposed exact formula, and secondly because it confirms
the reliability of standard canonical methods in the
quantization of quantum field theory.

The matrix element of a product of $n-2$ exponentials on
the cylinder is related to the correlator of $n$ exponentials
on a sphere by associating the states with two points 
on the sphere represented on the complex plane by, 
say $x_1=0$ and $x_n=\infty$.
Putting $\alpha_1=Q/2+iP$ and $\alpha_n=Q/2+iP^\prime$,
with $P,P^\prime$ real, the energies of the two states participating in the
matrix element are given by $E=2P^2$ and $E^\prime=2P^{\prime2}$.
Then the matrix elements calculated in Ref.~\cite{braatencgt}
were those of the operator $e^{2b\sigma\phi}$ 
between states with $P=bk$ and $P^\prime=bk^\prime$, and
they were expanded
in an asymptotic series in the limit $b\to0$. In the following section we
first develop the special function $\Upsilon(b\sigma)$,
which figures in the proposed formula for the exact three point function, 
in an asymptotic series for $b\to0$ with $\sigma$ fixed.
Using this result in the exact formula, we then expand
the proposed three point function to $O(b^{10})$
and find complete agreement with the results of \cite{braatencgt}.

\section{The Comparison}
\subsection{The Upsilon Function}
The construction of the Liouville three point function
\cite{dorno,zam1995} (for a recent review of Liouville Theory,
see \cite{teschner})
employs the the special function $\Upsilon(x)$
which satisfies the functional relations
\bea
\Upsilon(x+b)&=&{\Gamma(bx)\over\Gamma(1-bx)}b^{1-2bx}\Upsilon(x)\\
\Upsilon(x+1/b)&=&{\Gamma(x/b)\over\Gamma(1-x/b)}b^{-1+2x/b}\Upsilon(x).
\eea
To analyze the limit $b\to0$ define $g(x)=b^{x^2-bx}\Upsilon(x)$.
Then it is appropriate to focus on the first of these relations
which becomes in terms of $g$
\bea
g(x+b)={b\Gamma(bx)\over\Gamma(1-bx)}g(x).
\eea
First examine the $b\to0$ limit with $\sigma=x/b$ fixed. Define
$f(\sigma)=b^\sigma\Gamma(\sigma)g(b\sigma)$ so we have
\bea
f(\sigma+1)&=&{\Gamma(1+b^2\sigma)\over\Gamma(1-b^2\sigma)}f(\sigma)\\
&=&f(\sigma)\exp\left\{-2\gamma b^2\sigma-2\sum_{n=1}^\infty{\zeta(2n+1)\over2n+1}
(b^2\sigma)^{2n+1}\right\}.
\eea
We can develop a solution of this equation in perturbation theory
by making the {\it ansatz}
\bea
f(\sigma)=f_0\exp\left\{b^2\gamma\phi_1(\sigma)+
\sum_{n=1}^\infty b^{2(2n+1)}{\zeta(2n+1)\over2n+1}\phi_{2n+1}(\sigma)\right\}
\label{fansatz}
\eea
with $\phi_n(\sigma)$ a polynomial of order $n+1$, vanishing at
$\sigma=0$ and satisfying the relation
\bea
\phi_n(\sigma+1)&=&\phi_n(\sigma)-2\sigma^n
\label{phidefs}
\eea
A generating function for the $\phi$'s is
\bea
\sum_{n=0}^\infty {\eta^n\over n!}\phi_n(\sigma)\equiv-2{e^{\eta\sigma}-1
\over e^\eta-1}.
\eea
Since this generating function has a finite radius of convergence,
determined by the singularities closest to the origin ($\eta=\pm2\pi i$),
it follows that the $\phi_n(\sigma)$ grow like $n!$ so the perturbation
series diverges for finite $b$ and is only an asymptotic series.
We list the first 3 $\phi_{2n+1}$:
\bea
\phi_1(\sigma)&=&\sigma(1-\sigma)   \\
\phi_3(\sigma)&=&-{1\over2}\sigma^2(1-\sigma)^2   \\
\phi_5(\sigma)&=& -{1\over6}\sigma^2(1-\sigma)^2(2\sigma^2-2\sigma-1)  
\eea
Summarizing, we have obtained the asymptotic expansion, for $b\to0$
and $\sigma$ fixed:
\bea
\Upsilon(b\sigma)=b\Upsilon_0
{b^{-b^2\sigma^2+(b^2-1)\sigma}\over\Gamma(\sigma)}
\exp\left\{b^2\gamma\phi_1(\sigma)+
\sum_{n=1}^\infty b^{2(2n+1)}{\zeta(2n+1)\over2n+1}\phi_{2n+1}(\sigma)\right\}
\label{weakups}
\eea
where we have determined $f_0$ in terms of $\Upsilon_0\equiv\Upsilon^\prime(0)$
via $f_0=b\Upsilon_0$

The conditions so far imposed to get this result would also
be satisfied if (\ref{fansatz}) and hence (\ref{weakups})
were multiplied by a periodic function $X(\sigma)=X(\sigma+1)$, 
with $X(0)=1$\footnote{We thank J. Teschner for raising this
issue}. 
To show that in fact $X(\sigma)\equiv1$, we refer
to the theory of double gamma functions \cite{barnes}, 
which defines them by 
\bea
{d^3\over dz^3}\ln\Gamma_2(z|\omega_1,\omega_2)&\equiv&-2\sum_{m,n=0}^\infty
{1\over(z+m\omega_1+n\omega_2)^3}\nonumber\\
&=&{d^3\over dz^3}\ln\Gamma(z/\omega_1)
-\int_0^\infty dt {t^2e^{-zt}\over(e^{\omega_2t}-1)(1-e^{-\omega_1t})}.
\label{gamma2def}
\eea
The function $\Upsilon$ is defined in terms of 
$\Gamma_2(z|b,1/b)$. Clearly, this equation fixes $\ln\Gamma_2$
only up to an additive quadratic polynomial in $z$, which
is fixed by the functional relations $\Gamma_2$ must satisfy. Nevertheless,
setting $\omega_1=b=1/\omega_2$, the right side of (\ref{gamma2def})
can be developed in an expansion for small $b^2>0$ which can
be seen to be  consistent with the detailed
{\it ansatz} (\ref{fansatz}), up to an undetermined 
additive quadratic polynomial in the exponent. Since this {\it ansatz}
led uniquely to (\ref{weakups}), we conclude that
$X(\sigma)=\exp(\alpha\sigma^2+\beta\sigma)$.
Periodicity in $\sigma$ with period $1$ implies then that $\alpha=0$ 
and $\beta=2\pi Ni$ for
some integer $N$. But since $\Upsilon$ is a real analytic function,
we conclude that in fact $N=0$ so (\ref{weakups}) is indeed the
correct asymptotic expansion for small $b$.
\subsection{Formula for the Liouville Three Point Function}
Ref.~\cite{dorno,zam1995} propose the following three point
function for Liouville quantum field theory:
\bea
C(\alpha_1,\alpha_2,\alpha_3)&=&\left[\pi\mu b^{-2b^2}
{\Gamma(1+b^2)\over\Gamma(1-b^2)}
\right]^{(Q-\sum\alpha_i)/b}\nonumber\\
&&\hskip-.5cm\times{\Upsilon_0\Upsilon(2\alpha_1)\Upsilon(2\alpha_2)
\Upsilon(2\alpha_3)\over
\Upsilon(\alpha_1+\alpha_2+\alpha_3-Q)\Upsilon(\alpha_1+\alpha_2-\alpha_3)
\Upsilon(-\alpha_1+\alpha_2+\alpha_3)\Upsilon(\alpha_1-\alpha_2+\alpha_3)}
\hskip.25cm
\eea
We would like to compare this formula with the perturbation
expansion of Liouville on a cylinder given by Braaten, Curtright,
Ghandour, and Thorn \cite{braatencgt}. For that purpose two
of the $\alpha$'s, say 1 and 3, must be of the form $iP+Q/2$ with $P$ real.
These would represent the states at $t=\pm\infty$.
So write $\alpha_1=ibk+Q/2$ and $\alpha_3=ibk^\prime+Q/2$. The third
one is taken to be of $O(b)$: $\alpha_2=b\sigma$. Then the
arguments of all of the $\Upsilon$'s are small after exploiting
the identity $\Upsilon(x)=\Upsilon(Q-x)$. We have
\bea
2\alpha_1=Q+2ibk\to-2ibk,\qquad 2\alpha_2&=&2b\sigma,\qquad
2\alpha_3= Q+2ibk^\prime\to-2ibk^\prime\\
\alpha_1+\alpha_2+\alpha_3-Q&=&ib(k+k^\prime)+b\sigma\\
-\alpha_1+\alpha_2+\alpha_3&=&ib(k^\prime-k)+b\sigma\\
\alpha_1+\alpha_2-\alpha_3&=&-ib(k^\prime-k)+b\sigma\\
\alpha_1-\alpha_2+\alpha_3&=&Q+ib(k^\prime+k)-b\sigma\to
-ib(k^\prime+k)+b\sigma
\eea
Inserting these results into the formula gives
\bea
C(\alpha_1,\alpha_2,\alpha_3)&=&\left[\pi\mu b^{-2b^2}
{\Gamma(1+b^2)\over\Gamma(1-b^2)}
\right]^{-\sigma-ik-ik^\prime}\nonumber\\
&&\hskip-.5cm\times{\Upsilon_0\Upsilon(-2ibk)\Upsilon(2b\sigma)\Upsilon(-2ibk^\prime)
\over\Upsilon(b(\sigma+ik+ik^\prime))\Upsilon(b(\sigma-ik-ik^\prime))
\Upsilon(b(\sigma+ik-ik^\prime))\Upsilon(b(\sigma-ik+ik^\prime))}
\label{ourform}
\eea
Let us note that a factor of the form $e^{\xi x^2}$ contributing to
$\Upsilon(x)$ will always cancel in the three point formula
between numerator and denominator. This is simply because
\bea
(\sigma+ik+ik^\prime)^2+(\sigma-ik-ik^\prime)^2+
(\sigma+ik-ik^\prime)^2+(\sigma-ik+ik^\prime)^2
=(2\sigma)^2+(-2ik)^2+(-2ik^\prime)^2, \nonumber
\eea 
the cross terms all canceling. On the other hand, a factor of the
form $e^{\xi x}$ does not cancel between numerator and denominator,
and it yields a net factor $e^{b\xi(-2ik-2ik^\prime-2\sigma)}$.

We now substitute Eq.~\ref{weakups} in the Eq.~\ref{ourform}.
Gathering the contributions of the factors multiplying the 
exponential in Eq.~\ref{weakups} and setting the $b^2=0$
in the gamma functions in the square
brackets leads to the zeroth 
approximation 
\bea
C_0(k,\sigma,k^\prime)&=&{1\over b}\left[\pi\mu b^{-2b^2}
\right]^{-\sigma-ik-ik^\prime}b^{-2(b^2-1)(\sigma+ik+ik^\prime)}\nonumber\\
&&\hskip-.5cm\times{\Gamma(\sigma+ik+ik^\prime)\Gamma(\sigma-ik-ik^\prime)
\Gamma(\sigma+ik-ik^\prime)\Gamma(\sigma-ik+ik^\prime)
\over\Gamma(-2ik)\Gamma(2\sigma)
\Gamma(-2ik^\prime)
}\\
&=&{1\over b}\left[{b^2\over\pi\mu}
\right]^{\sigma+ik+ik^\prime}
{|\Gamma(\sigma+ik+ik^\prime)
\Gamma(\sigma+ik-ik^\prime)|^2
\over\Gamma(-2ik)\Gamma(2\sigma)
\Gamma(-2ik^\prime)
}
\label{zeroform}
\eea
This is to be compared to Eq.~(12) of Ref.~\cite{braatencgt}
after translating the parameters of that work to the
ones used here:
\bea
g=b\sqrt{2\pi},\qquad m=b\sqrt{\mu\pi},\qquad \alpha=2\sigma,
\qquad k^\prime ,k^{\prime\prime}\to 2k,2k^\prime\ .
\eea
In this dictionary quantities on the left are those of  Ref.~\cite{braatencgt}
and those on the right are those of the current work. 
Then the formula from the older work becomes
\bea
\bra{2k^\prime}e^{2g\sigma q}\ket{2k}
&=&{1\over 2(2\pi)^{3/2}b}\left[{b^2\over\pi\mu}
\right]^{\sigma}
{|\Gamma(\sigma+ik+ik^\prime)
\Gamma(\sigma+ik-ik^\prime)|^2
\over\Gamma(2\sigma)|\Gamma(-2ik)\Gamma(-2ik^\prime)|}.
\label{oldzeroform}
\eea
We see that the two formulae agree up to a phase redefinition of the
initial and final states and an overall normalization constant
independent of the states, operator, and coupling constant.

Next we turn to the perturbative corrections to the zeroth order
result. These are contained in the ratio of gamma functions
inside the square brackets as well as the contribution of the
exponentials in Eq.~\ref{weakups} to the ratio of upsilon functions
in (\ref{ourform}). Define
\bea
\Phi_{2n+1}&\equiv&\phi_{2n+1}(2\sigma)+\phi_{2n+1}(-2ik)
+\phi_{2n+1}(-2ik^\prime)\nonumber\\
&&-2{\rm Re}\ \phi_{2n+1}(\sigma+ik+ik^\prime)
-2{\rm Re}\ \phi_{2n+1}(\sigma+ik-ik^\prime).
\eea
Then the perturbative corrections are contained in a factor
\bea
F&=&\left[{\Gamma(1+b^2)\over\Gamma(1-b^2)}\right]^{-\sigma-ik-ik^\prime}
\exp\left\{b^2\gamma\Phi_1
+\sum_{n=1}^\infty b^{4n+2}{\zeta(2n+1)\over2n+1}\Phi_{2n+1}\right\}\\
&=&\exp\left\{b^2\gamma(\Phi_1+2(\sigma+ik+ik^\prime))
+\sum_{n=1}^\infty b^{4n+2}{\zeta(2n+1)\over2n+1}
\left(\Phi_{2n+1}+2(\sigma+ik+ik^\prime)
\right)\right\}
\eea
In the weak coupling calculations of \cite{braatencgt} the
factor $F$ was evaluated through order $O(g^{10})$, and it is
of interest to compare those results with the exact
proposal. So we work out the first three $\Phi$'s:
\bea
\Phi_1&=& -2ik-2ik^\prime-2\sigma\\
\Phi_3&=& -6(k^2-k^{\prime2})^2-12\sigma(\sigma-1)
(k^2+k^{\prime2})-6\sigma^4+{4}\sigma^3
+i{8}(k^3+k^{\prime3})\\
\Phi_5&=&20(k^2-k^{\prime2})^2(k^2+k^{\prime2})-10(k^2-k^{\prime2})^2
-20\sigma^6+28\sigma^5-10\sigma^4-32ik^5-32ik^{\prime5}\nonumber\\
&&-20\sigma^2(1-\sigma)^2(k^2+k^{\prime2})
-20\sigma(1-\sigma)(k^4+6k^2k^{\prime2}+k^{\prime4})
\eea
We see that the first term in the exponent cancels so that $F$
simplifies to
\bea
F=\exp\left\{\sum_{n=1}^\infty b^{4n+2}{\zeta(2n+1)\over2n+1}
\left(\Phi_{2n+1}+2(\sigma+ik+ik^\prime)
\right)\right\}.
\eea
Thus we immediately confirm the structure found in \cite{braatencgt}
that through $O(b^{10})$ the only nonvanishing terms are $O(b^6)$
and $O(b^{10})$. In fact, after translating the parameters
from the older work to the ones in the current work,
the values of those terms agree
exactly with what we have found here.

The weak coupling calculational program given in \cite{braatencgt}
can of course be carried out to any desired order. For the
three point function we have the exact answer so perturbation
theory gives no additional information. However the main
virtue of perturbation theory is that it can be applied
to an arbitrarily complicated process. Thus it can
give complementary information to bootstrap approaches.

\vskip.5cm
\noindent\underline{Acknowledgments:}
I am grateful to J. Teschner for
stimulating my interest in this exercise and for
very helpful comments and discussions. I also thank
N. Seiberg for helpful comments on the manuscript.
This work was supported in
part by the Monell Foundation and in part by
the Department of Energy under Grant No. DE-FG02-97ER-41029.


\end{document}